# Advanced Characterization Methods for Crystals of an Extraordinary Material

Vivekananda Bal[≠], Jackie M. Wolfrum[§], Paul W. Barone[§], Stacy L. Springs[§],

Anthony J. Sinskey[§ζ], Robert M. Kotin[§]*, and Richard D. Braatz[≠§]

[≠]Department of Chemical Engineering, Massachusetts Institute of Technology, Cambridge, MA, USA

[§]Center for Biomedical Innovation, Massachusetts Institute of Technology, Cambridge, MA, USA

[ζ]Department of Biology, Massachusetts Institute of Technology, Cambridge, MA, USA

* Department of Genetic & Cellular Medicine, University of Massachusetts Chan Medical School,

Worcester, MA, USA

**Abstract:** Physicochemical characterization of materials is central to the field of science and engineering and is essential to design and develop new materials with altered activities or specific properties. Many of the assays that are available for small molecules, e.g., XRD, NMR LC-MS, cannot be applied to large, macromolecular species easily and many of these techniques are tedious. Thus, it is extremely challenging to characterize complex materials such as crystals of adeno-associated virus (AAV) capsids (molecular weight 5.8 MDa) with or without DNA genomes. This article characterizes rAAV serotypes 5, 8, and 9 capsids as model biological macromolecular assemblies. Capsid crystals produced by hanging-drop vapor diffusion are characterized in situ using cross-polarized light microscopy and ex-situ using scanning electron microscopy (SEM) imaging, energy dispersive X-ray spectroscopy (EDX) analysis, and transmission electron microscopy (TEM) selected area electron diffraction (SAED). Cross-polarized light can be used to identify capsid crystals within a heterogenous system containing crystals of kosmotropic or chaotropic salts, fibers as impurities, and dense solid phase or opaques crystals, which commonly form during crystallization of massive proteinaceous assemblies. Despite highly conserved structures among the AAV capsids, the crystal birefringence and optical retardation suggests that crystalline capsids possess AAV serotype-specific structural or electronic environment differences. SEM imaging demonstrated that the crystal growth occurs by random 2D nucleation forming islets followed by the growth of 2D islets by kink-site attachment and/or formation of more 2D nuclei to spread the already formed islets. Biological macromolecular assemblies also form semi-crystalline solids, that appear as dense solids or opaque materials, as suggested by the presence of crystallites in amorphous phase. Elemental analysis using EDX shows that C, O, and N are the most abundant elements in the ratio of $2.33 \pm 0.222 : 1 : 0.583 \pm 0.019$ for rAAV serotypes 5, 8, and 9 and can be an alternative to the conventional protein sequencing for AAV (and other virus) identification. Biological macromolecular assemblies such as capsids with ~4152 atoms of low atomic number and numerous covalent bonds are found to facilitate the plural scattering responsible for the Kikuchi diffraction patter in electron diffraction even for thin crystals in the range of ~300 nm to ~900 nm thickness (at 400 keV accelerating voltage). For an optimal spot diffraction pattern, crystals must possess at least one dimension consisting of at most 8 layers of capsids. It is found that capsids form body-centered orthorhombic unit cells with four full complete capsids and tetragonal unit cells with two complete capsids.

## 1. Introduction

Analysis of solid forms of a material is extremely important in understanding structure, composition, mechanical and electrical properties, feasibility of long-term storage and transport, and the dissolution characteristics.[1–5] Specifically, for biological materials, understanding their structure, composition, and dissolution characteristics may help in predicting biological activity, solubility, thermal stability, etc. and in identifying the surface specific region responsible for the specific functions such as bindings to the cells or antibodies or antigens.[6–10] In addition, detailed structural and composition analysis may provide insight into developing engineered biomaterials as a means of altering biological activity for specific applications.[9,11] Characterization of materials in the solid phase is routine for small inorganic and organic molecules.[5,12–16] Likewise, low-molecular weight biological molecules such as protein molecules are routinely crystallized and often characterized to understand how these proteins interact with other molecules that determines the mechanisms of action which are critical for developing new drugs or therapies that interact with these proteins.[11,17–20] In addition, the protein crystal properties may be useful in determining the stability for long-term storage and transport in crystal form, and the dissolution properties of the crystals for reformulation etc.[21] For most of these proteins, the crystal quality and structure are determined using the "gold standard" X-ray diffraction, or electron diffraction analysis, and a less frequently, using cross-polarized light microscopy.[11,17,20,22–24] For larger biological molecules, such as mRNa (200–500 kDa), and mABs (~ 150 kDa), and for massive macromolecular assemblies such as capsids (~ 3.6 MDa), no reports are available that characterize the crystal parameters beyond the crystallographic structures.[6,8,9,25,26] AAV2 was the first AAV serotype determined by X-ray crystallography.[9] Subsequently, numerous AAV serotypes and capsid variant structures have been determined by X-ray diffraction.[6,8,25] Subsequently, the first cryo-EM structure of AAV was determined at a comparable resolution of 2.8 Å in 2015 and has become the method of choice for AAV capsid structural analysis.[27,28] However, achieving high resolution of ~ 1–1.5 Å, is challenging for cryo-EM, but routinely achieved in X-ray diffraction.[28] This high-resolution structural analysis for more-complex structures e.g., capsid/receptors or capsid/mAb, provides information necessary for engineering the capsids to gain or lose specific functions, e.g., cell-surface receptors or Ab recognition, respectively.[9,28] For complex structures, such as capsids, high-energy synchrotrons were the source of the X-rays, which are not accessible as a routine laboratory analytical tool.[9] This necessitates the use of an easily accessible methods, which can provide high resolution and are able to analyse the less ordered crystals without damaging the sample.

Thus, the above discussion suggests that, for biological macromolecular assemblies, starting from the in-situ screening of the crystal phase to the mechanism of crystal formation, crystal composition, and the structural irregularities have not been explored and there is a need of assessing the potential of simple and easily accessible techniques, which are regularly used in material science. This work characterizes crystals of the capsids of adeno-associated virus serotypes 5, 8, and 9 as a model biological macromolecular assemblies using cross-polarized light microscopy for screening of crystal phase, SEM imaging for studying crystal formation and structural information, EDX for analyzing composition, and SAED (selected area electron diffraction) for understanding the crystal structure.

AAV vectors (recombinant, rAAV) are commonly used in research and for gene therapy applications to deliver DNA to somatic cells that are deficient for certain gene products, typically caused by mutations.[29] The AAV particle, also known as capsid, has icosahedral symmetry (T = 1) composed of 60 virion protein (VP) subunits, VP1 (87 kDa), VP2 (72 kDa), and VP3 (62 kDa) in approximately 1:1:10 ratio.[30] There at least 11 serotypes of AAVs (1 through 11) as well as numerous naturally occurring and artificially derived capsid variants that have been used to produce rAAV. Like the virus, the vectors also contain a linear, single-stranded DNA genome, approximately 4.7kb. However, the vector genome is devoid of AAV genes retaining only the non-coding, cis-acting inverted terminal repeats (ITRs). Three serotypes of rAAV, derived from AAV5, AAV8, and AAV9, were selected for study. The corresponding capsid models, are shown in **Fig. 1a,** where capsids are radially colored by distance from the center (as obtained from the VIPER database,[31] https://viperdb.org) to indicate the location of proteins. It shows some protrusion as well as depression regions on the capsid surface. It is difficult to understand the slight differences existing between three serotypes by simply looking at the models of intact capsids as shown in **Fig. 1a.** VP3 ribbon and mesh surface models in **Fig. 1b** shows the subtle differences between the three serotypes. Their characterization will help not only in designing new drugs, understanding structure or fundamentals, but also in developing a separation/purification process based on crystallization and in producing diffraction-grade crystals.[32]

Detailed structural and composition analysis of crystals of proteinaceous assemblies such as capsids will help in engineering the capsids by identifying the neutralizing antibody "hotspot" regions (or dominant epitopes) on the capsid surface, mapping of receptor interaction sites, location of capsid pentameric and hexameric subunits and potential opening for genome release and identifying how it enters the host cells.[9,28,29] Additionally, capsid crystallization as a downstream processing step may result in stabilization for long-term storage and transport methods, understanding solubility, thermal stability, and dissolution characteristics. Particularly for complex biotherapeutics such as capsids, in situ screening of the crystal phase is challenging because of the tendency to form dense solid phase, opaque material, and many other unidentified structures. These amorphous assemblies are not seen in the crystallization of low-molecular weight proteins or biologics, and other proteins/biological impurity molecules and salt material.[33] Cross-polarized light microscopy, regularly used in optical minerology and the gem industry, can be a suitable method for rapid in-situ screening in case of a large number of experimental conditions[34,35] The large size of capsids (25 nm diameter) make them clearly visible in SEM. This enables the study of the crystal phase growth mechanism and capsids arrangement in crystals, and crystal structural irregularities in SEM feasible, which measurements are commonly performed using AFM (atomic force microscope) for the growth of nanoscale metal/metal oxides/proteins.[36,37] Sequencing of amino acids in a biological molecule can provide an idea of elemental composition in that molecule. Sequencing of a heavy biological macromolecular assembly such as a capsid, however, is a difficult and tedious process. EDX analysis commonly used in material science for characterization of thin films of inorganic or small organic molecules is a rather simple and fast method and therefore, can be a suitable alternative to the sequencing analysis.[10,38] Additionally, capsids have massive, complex structures with excessively large number of solvent contacts that reduce electron density and weaken or diffuse the diffraction signal. This significantly increases the difficulty of X-ray diffraction and needs highly ordered diffraction grade crystals and the use of high-energy X-ray source such as a synchrotron source. Developing electron diffraction methods for capsid structural analysis, which provides relatively higher resolution of < 1 Å in most cases than the gold standard X-ray diffraction (max 1 Å) and the cryo-EM (2–2.5 Å),[28,39] would increase the opportunity for researchers to conduct these studies in their own laboratories or institutions without the need for a particle accelerator (synchrotron).

Although this work uses rAAV capsids as model biological molecular assemblies, the characterization methods presented in this work can also be applicable to the characterization of mABs, mRNAs, and other small molecules and macromolecular assemblies.

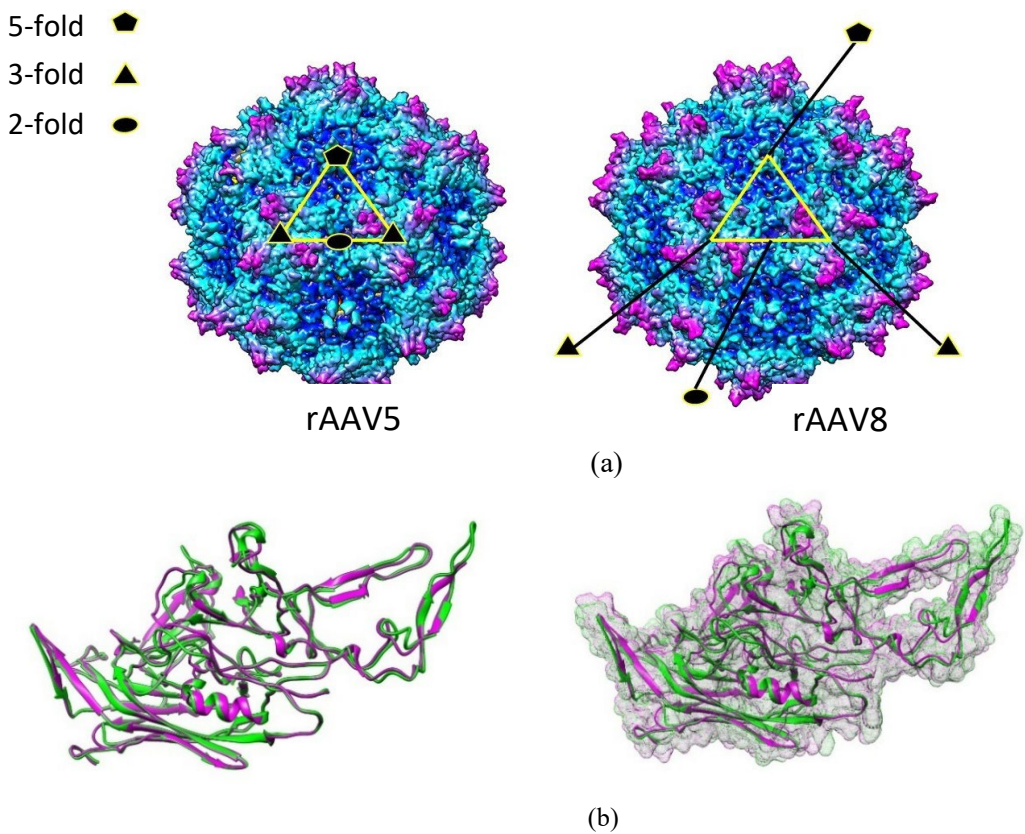

**Fig. 1:** (a) AAV5 and AAV8 capsid models produced using UCSF Chimera. The capsids are rendered at 2 nm resolution and are radially colored by the radial distance from the center (red 9 nm, yellow, 10 nm, dark blue 11 nm, cyan 12 nm, violet 14 nm). The atomic coordinates were obtained from the VIPER database[31] (https://viperdb.org). Models show the 2- , 3- , and 5- fold symmetry axis for capsid as indicated by the circle, triangle, and pentagon, respectively. (b) The ribbon (left) and mesh (right) surface models of the VP3 subunits of AAV5 (magenta) and AAV8 (green) capsids.

## 2.1 Materials

Chemicals used in the experiments were of molecular biology grade: Sodium dihydrogen phosphate dihydrate (Bio Ultra, ≥99.0%, SigmaAldrich, St Louis, MO, USA), 1M hydrochloric acid (Bioreagent, SigmaAldrich, St Louis, MO, USA), sodium hydroxide (BioXtra, ≥98% anhydrous, SigmaAldrich, St Louis, MO, USA), sodium chloride (BioXtra, ≥99.5%, SigmaAldrich, St. Louis, MO, USA), potassium dihydrogen phosphate (Bio Ultra, ≥99.5%, SigmaAldrich, St. Louis, MO, USA), potassium chloride (BioXtra, ≥99.5%, SigmaAldrich, St. Louis, MO, USA), polyethylene glycol (PEG8000, Bio Ultra, SigmaAldrich, St Louis, MO, USA), phosphate buffer saline (PBS 1X (150 mM sodium phosphate and 150 mM NaCl), pH 7.2, Bio Ultra solution, SigmaAldrich, St Louis, MO, USA), phosphotungstic acid (10% solution, SigmaAldrich, St Louis, MO, USA), Poloxamer-188 (Pluronic F-68, 10%, Bioreagent, Thermofisher, Waltham, MA, USA), Copper grid (carbon support film, 200 mesh, Cu; CF200-CU-50, EMS, Hatfield, PA, USA) and Glutaraldehyde solution (10% aqueous solution, EMS, Hatfield, PA, USA).

## 2.2 rAAV samples

Full capsids (capsids with vector genome, vg) of the recombinant adeno-associated virus serotypes 5 (rAA5), 8 (rAAV8), and 9 (rAAV9) were purchased from Virovek at a concentration of $10^{14}$ vg/ml and pH 7.2 in PBS buffer. The samples designated as "full rAAV5", "full rAAV8", and "full rAAV9" actually contained 80% full and 20% empty capsids. The full capsids carried a genome of length 2.46 kbp (Virovek) with cytomegalovirus (CMV) E1a enhancer promoter and an open reading frame that encodes green fluorescent protein (GFP). Samples from Virovek were aliquoted into four equal parts and stored at −80°C for long-term use. For immediate use, a small vial was stored at 4°C, at which AAV is stable for at least 4 weeks.

## 2.3 Experiment

Hanging-drop vapor diffusion experiment was conducted in a 24-well crystallization plate (VDX, Hampton Research, California, USA) to crystallize capsids (**Fig. 2**).[40,41] Each well contains a suspended droplet hanging from a glass slide at the top of the well, and a reservoir solution at the bottom of the well. All wells contained a 2 μL droplet and 1 mL reservoir solution. Both the reservoir solution and suspended droplet contain salt (NaCl) and polyethylene glycol (PEG; molecular weight 8000) as precipitants in a phosphate-buffered saline (PBS) solution, with AAV only in the droplet. Each droplet is produced as a 1:1 mixture of rAAV sample in PBS (1 μL) and reservoir solution (1 μL) resulting in the initial concentration of PEG and salt in the droplet to half of that in the reservoir solution. Once each well is covered, the vapor pressure depression resulting from the higher salt concentration in the reservoir results in water vapor diffusion from the droplet to the reservoir as described in our previous work.[42] This causes the rAAV concentration in the droplet to increase and eventually, quasi-steady state is attained (**Fig. 2**). Once supersaturation condition is achieved, rAAV capsid crystal nucleation begins and crystal growth follows.

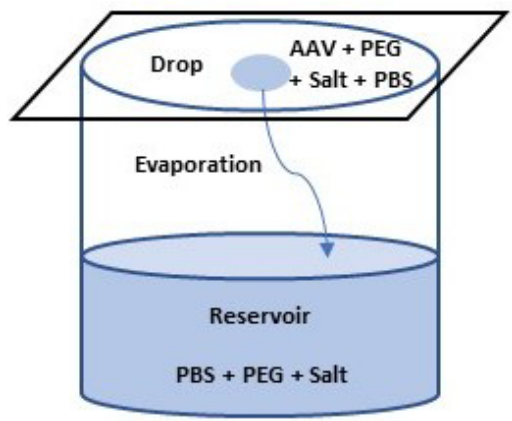

**Fig. 2:** Hanging-drop vapor diffusion experimental setup.

Each droplet was monitored using optical microscopy (Imaging Source DMK42BUC03) at regular time intervals to examine the progress and track the evolution of crystals/particles formation. Cross-polarized light microscopy (Leica Z16 APO) was used to confirm the crystallinity of the particles. The pH of the solutions used in the experiments is well above pH 5.5 since capsid stability may be compromised at more acidic conditions,[43,44] Crystallization experiments were performed at room temperature, (23±2°C).

### 2. 3. 1 Cross-polarized light microscopy

Cross-polarized light microscopic analysis was performed in a Leica Z16 APO microscope. For the study, a 24-well plate with a target well/drop was positioned on the microscope base. Then the white light is passed through a polarizer at the base of microscope and the polarizer produces plane polarized light, which is then allowed to pass through the crystals. Then the analyzer, which is also a polarizer and placed at the top, is rotated 360° gradually, while keeping the crystal fixed in position. Upon rotation of the analyzer, optically-anisotropic crystals produce a spectrum of interference color at all angular positions except at the 90°, 180°, 270°, and 360° positions with respect to the base polarizer, where interference is destructive and the crystals appear dark/extinct. This phenomenon, well known as a birefringence, confirms the anisotropic nature of the crystals. This work describes the procedure to use the cross-polarized light to identify/screen crystals of capsids from a mixture of impurity solids such as hair fibers and other solid forms such as semisolids, which are common in the crystallization of macromolecular proteinaceous assemblies, yet are not discussed in the literature.

### 2.3.2 Cryo-electron microscopic analysis

Crystallinity of the particles was further verified and additional structural information was obtained using electron diffraction in cryo-TEM.[45,46] For cryo-TEM imaging, a 1:1 mixture (6 μL total) of reservoir solution and 1X PBS buffer of pH 7.2 is added into a droplet containing well developed crystals with sharp edges and then crystals were detached from the glass surface using micropipette tip under continuous microscopy monitoring. Crystals were then broken into small pieces as small as possible using micropipette tip and the capsids in the crystal fragments were pre-fixed by adding 2.5% aqueous glutaraldehyde solution to inactivate them. For cryo-TEM sample preparation, a standard procedure as described in literature was followed.[22] Briefly, a 3-μL pre-fixed crystal fragments solution was dropped on a lacey copper grid having a continuous carbon film coating and blotted to remove excess sample by Cryo-Plunge III (Gatan) without damaging the carbon layer. The grid was then mounted on a single tilt cryo-holder (Gatan 626) equipped in the TEM chamber. Both the specimen and the holder tip were cooled down using liquid nitrogen. Imaging was done on a JEOL 2100 FEG microscope (Koch Institute MIT) using a minimum dose method, which is important to minimize damage of samples by the electron beam. The microscope was operated at 200–400 kV for collection of diffraction pattern. The diffraction patterns were recorded on a Gatan Orius SC200D (833).

### 2.3.3 SEM imaging

For the scanning electron microscope (SEM) imaging of the particles, droplets containing intact and well-developed particles with sharp edges were selected. Post-nucleation, growth of the crystals mostly continues for 1–2 weeks. For SEM imaging at higher magnification, crystals were collected approximately 72 h post-nucleation during the crystal growth phase. To collect the crystals, the droplet is diluted with 10 μL of 1:1 mixture of the reservoir solution and 1X PBS buffer of pH 7.2 to facilitate the manipulations. The crystals were then detached from the glass surface using micropipette tip, aspirated and dispensed on a centrifugal filter of 0.1 μm pore size (Millipore Sigma) and centrifuged at 2000g for 1 min to remove the solution and free capsids. Then the filter paper is cut out of the centrifuge tube and attached on a carbon tape coated brass stub. To make the surface conducting, a 10-nm thick gold coating (EMS Quorum, EMS 150T ES, MIT Material Science) is then applied on the crystal sample. Then imaging was performed in a high-resolution electron microscope (ZEISS merlin, MIT Material Science).

### 2.3.4 EDAX analysis

Sample preparation for energy dispersive X-ray (EDAX) analysis remains the same as that for SEM imaging except no gold coating is needed for EDAX analysis. EDAX analysis was performed in MIT material science division using an Octane Elect Super Detector attached with SEM microscope (ZEISS Merlin). As the crystals were lying on the membrane surface, the elemental mapping/analysis was performed first on the membrane itself as a control and then on the crystals, which also includes the underlying membrane surface. At the end, line EDAX were performed along the crystal axis. Energy of the X-ray beam was varied between 5–20 kV depending on the thickness of the crystals being analysed. In all the analysis, > 100 scans were used unless mentioned. The objective of the line EDX analysis performed in this work is to determine the relative presence of the different elements in a biological macromolecular assembly such as capsid.

## 3 Results and Discussion

This section describes in detail the method of screening of crystals in the droplet followed by the analysis of their growth mechanism, composition, and crystal structure.

This work describe how cross-polarized light can be used to discriminate crystals from other solid material impurities including the amorphous solids, semicrystalline materials, salt crystals, and contaminants such as hair fibers, which are commonly found in the crystallization medium. A wide range of crystallization conditions were screened in a hanging drop vapor diffusion experiment (**Fig. 2**) using normal microscope (Imaging Source DMK42BUC03) and cross-polarized light microscope (Leica Z16 APO).

**Fig. 3** shows the corresponding sample images of crystals. The crystals appear grey in normal microscope (**Fig. 3a1, b1, c1, d1, e1, f1**).

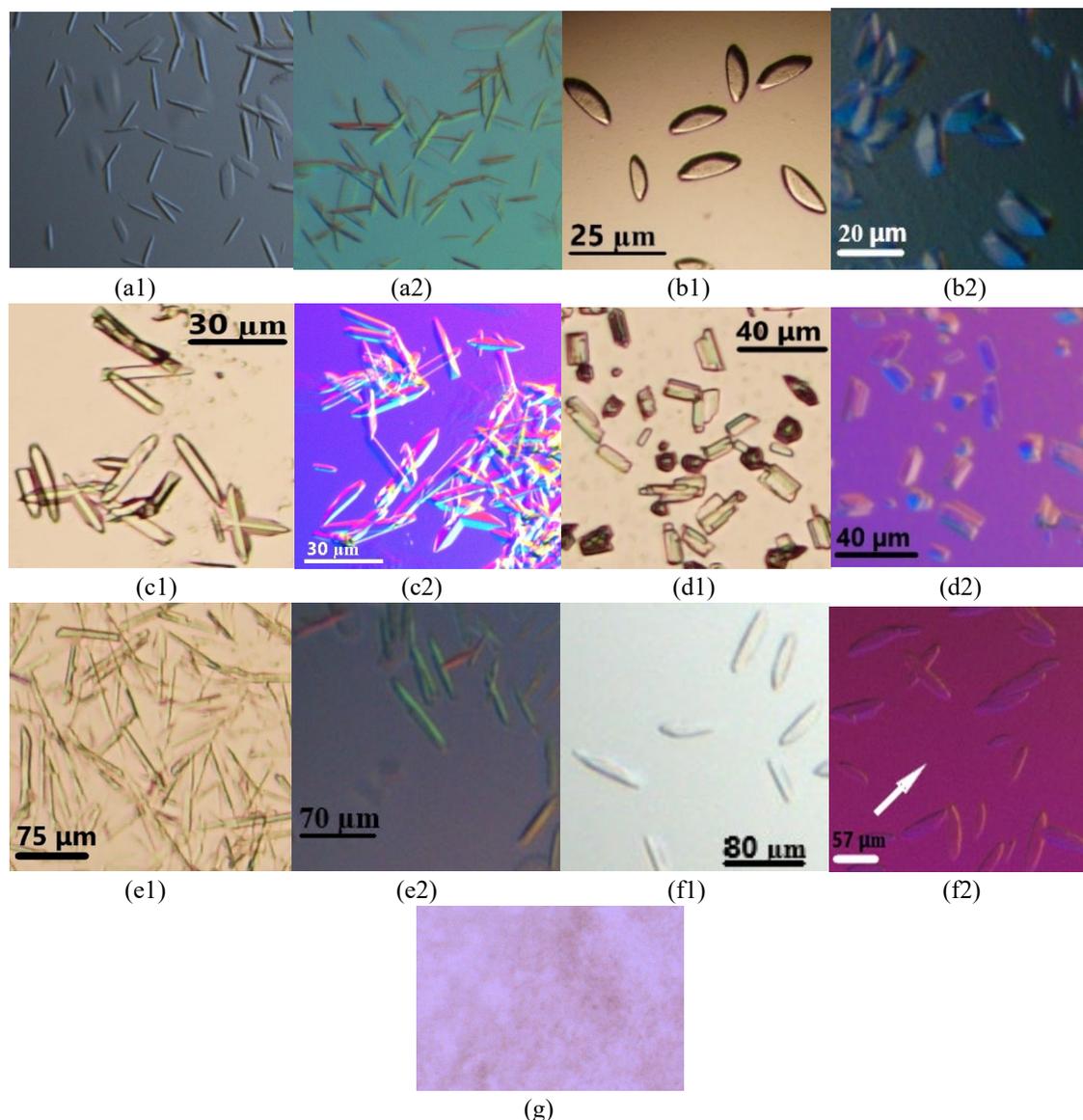

**Fig. 3:** Sample microscopic images of the crystals, showing (a) "empty" AAV5 crystals (1.5% PEG, 0.125 M NaCl, pH 5.7) under (a1) normal and (a2) polarized light, (b) "full" AAV5 crystals (2% PEG, 0.75 M NaCl, pH 6.1) under (b1) normal and (b2) polarized light, (c) "empty" AAV9 crystals (3% PEG, 0.3 M NaCl, pH 6.3) under (c1) normal and (c2) polarized light, (d) "full" AAV9 crystals (3.5% PEG, 0.5 M NaCl, pH 6.3) under (d1) normal and (d2) polarized light, (e) "empty" AAV8 crystals (3% PEG, 0.6 M NaCl, pH 5.7) under (e1) normal and (e2) polarized light, (f) "full" AAV8 crystals (3.5% PEG, 0.55 M NaCl, pH 5.7) under (f1) normal and (f2) polarized light, (g) image of precipitate under a polarized light microscope.

When anisotropic crystals are observed in cross-polarized light, crystals produce interference colors (**Fig. 3a2, b2, c2, d2, e2, f2**). In a cross-polarized light microscopy, a non-polarized white light is passed through a polarizer (filter) fixed in position to produce plane polarized light, which is passed through the crystal mounted on the microscope stage. For an anisotropic crystal, the incident plane-polarized light is split into two rays, one fast-moving ray (extra ordinary or e ray) and another slow-moving ray (ordinary or o ray) due to the direction-dependent speed of the light through the crystal (i.e., refractive index of the material depends on the polarization and propagation direction of light, birefractive). These two light wave rays are perpendicular to each other and the ray analyzer, which is also a polarizer and perpendicular to the first polarizer, combines these two rays and allows only those rays which are parallel to the analyzer and thus parallel to each other to pass. Since one wave is retarded with respect to the other, either constructive or destructive interference occurs between the split waves as they pass through the analyzer. Therefore, on rotation of analyzer with respect to the polarizer, crystals exhibit vibrant distinct interference color patterns (constructive interference) at every position between analyzer and polarizer, except at every 90° position (i.e., 90°, 180°, 270°, and 360°), where crystals remain extinct (as if vanished; e.g., diagonal crystal in **Fig. 4a, d, g, j**). At these positions, optical axis of a crystal aligns perfectly with the direction of polarization of the light coming from the polarizer and no interaction occurs between light and crystal and no light is able to pass through the analyzer (destructive interference).[47–49] This phenomenon, known as birefringence, is shown in the **Fig. 4** for rAAV5 full capsid crystals for different analyzer positions and is key to identification of anisotropic crystals. Particles obtained in the vapor diffusion experiment are birefringent and produce interference color as shown in **Fig. 3a2, b2, c2, d2, e2, f2**. This confirms that the particles are optically anisotropic.

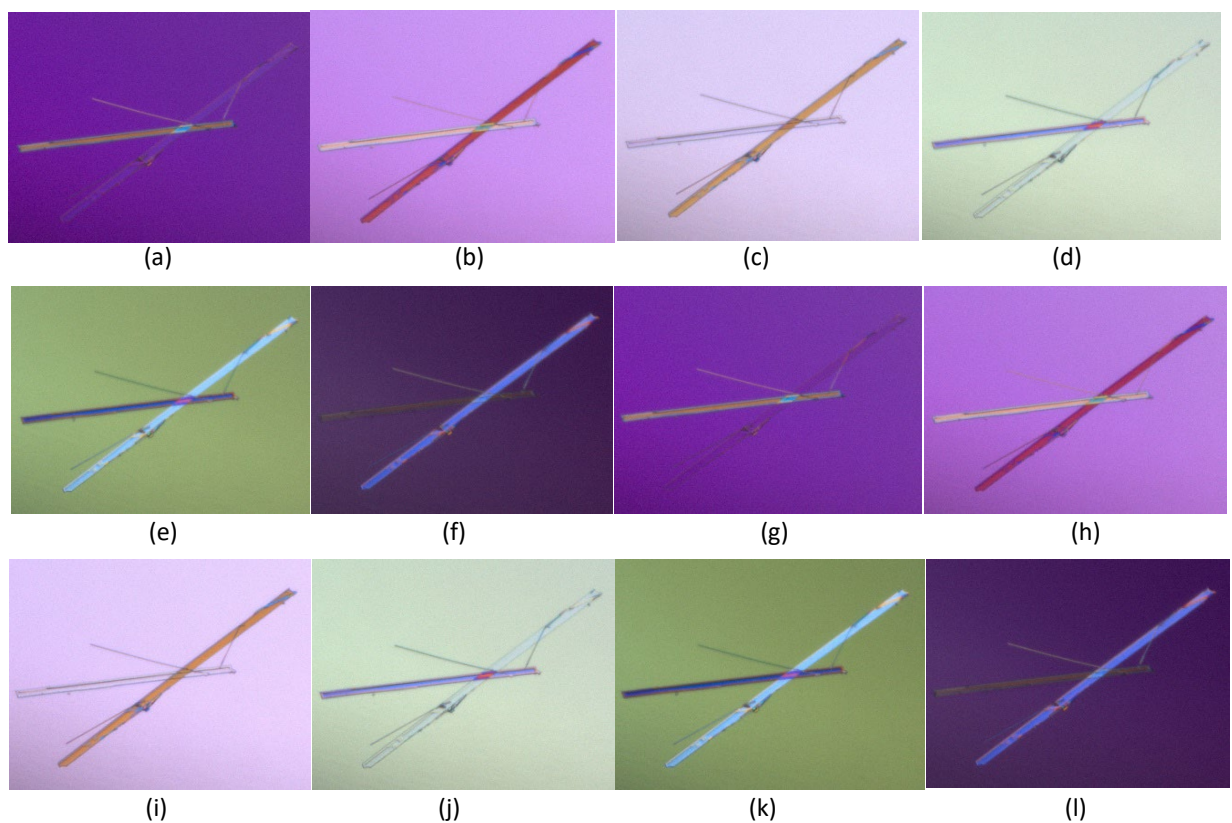

(a) (b) (c) (d)
(e) (f) (g) (h)
(i) (j) (k) (l)

**Fig. 4:** Images of materials in cross-polarized light microscope when the material is fixed on the base of microscope and analyzer is rotated at different angles. Images of crystals of full capsids of rAAV5 when the analyzer is rotated at an angle of $0°(a), 30°(b), 60°(c), 90°(d), 120°(e), 150°(f), 180°(g), 210°(h), 240°(i), 270°(j), 300°(k), 330°(l)$.

For isotropic crystals, the refractive index is same in all directions and therefore, the light coming from the polarizer is transmitted through the crystals and no interaction occurs between light and crystal and no light is able to pass through the analyzer. Thus, isotropic crystals remain extinct throughout a full rotation (0–360°) of the analyzer (**Fig. 5 row 1**). Thus, it's easy to distinguish anisotropic crystals of proteinaceous assemblies from the salt crystals as the later generally form cubic crystals. Crystals in **Fig. 3** are definitely single crystals as polycrystalline materials behave as isotropic materials (i.e., no birefringence property).[47,50]

Complex macromolecular assemblies tend to form optically dense solid materials/opaque crystals (**Fig. 5**

row 2) as observed in some of the experimental conditions for capsids as reported in our previous work.[33] These materials do not allow the plane polarized light from the polarizer to pass/transmit through them as they lack the internal structure required to interact with the polarized light and therefore, dense solids/opaque crystals will appear completely black against a bright background and not show birefringence in a cross-polarized light microscope regardless of the rotation angle of analyzer as shown in **Fig. 5 row 2**. However, reflected light microscopy is better for identification or analyzing color, texture or other physical properties of the opaque crystals.[51] Precipitates are easily identifiable visually by their morphology without the need of cross-polarized light microscope as shown in **Fig. 3g**.

Like anisotropic crystals, other solids such as hair materials, cotton fibers, which are very common in the crystallization medium as contaminants, also show birefringence/interference colors (**Fig. 5 row 3**) in cross-polarized light microscope, but their birefringence patterns differ. For example, crystals will show bright, vibrant interference colors and distinct interference patterns on rotation of the analyzer as show in **Fig. 4**, whereas hair fibers will exhibit a muted banded pattern and less pronounced birefringence (**Fig. 5 row 3**). Thus, birefringence can be used to screen crystals from other solid material impurities including the amorphous solids, semicrystalline materials, salt crystals, and contaminants such as hair fibers.

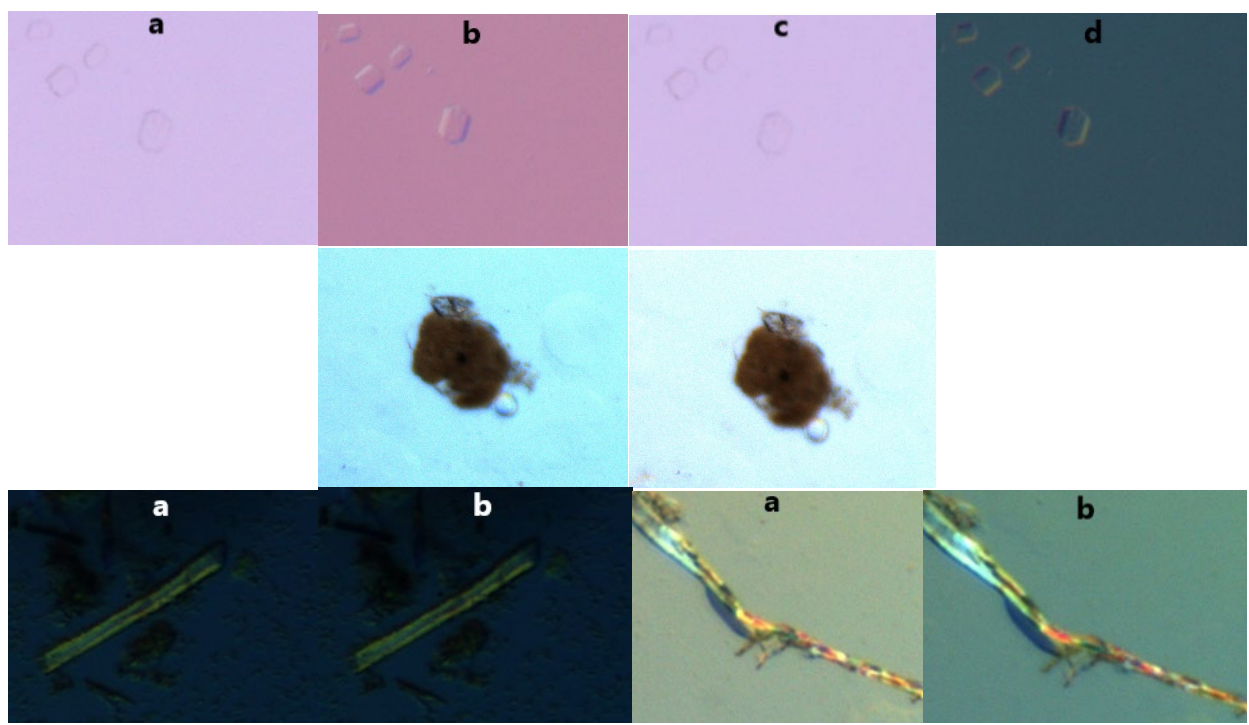

**Fig. 5:** Images of materials in cross-polarized light microscope when the material is fixed on the base of microscope and analyzer is rotated at different angles. Row 1: images of isotropic materials (salt crystals) when analyzer is rotated at an angle of $30°(a), 60°(b), 120°(c), 150°(d)$, Row 2: images of opaque materials/crystals when analyzer is rotated at an angle of $30°(a), 60°(b)$, and Row 3: images of hair fiber (left) or cotton (right), when analyzer is rotated at an angle of $30°(a), 60°(b)$.

The birefringence property of a crystal can also be used to obtain many of its physical properties. Crystal birefringence (i.e., the maximum difference in refractive index of the material, $|n_e - n_o|$, i.e., the difference in refractive indices of fast- and slow-moving rays) was calculated using Michel-Levy chart.[47] This calculation gives higher order birefringence of ~0.25 for (**Fig. 3a2**), and ~0.37 for

(**Fig. 3b2**). The retardation (i.e., the difference in optical path of fastest ray with respect to the slowest ray,[47,49] $\Gamma = d|n_e - n_o|$) is found to be ~630 nm for (**Fig. 3a2**), ~1120 nm for (**Fig. 3b2**). Similar birefringence (~0.45 for empty rAAV9, ~0.32 for full rAAV9, ~0.43 for empty rAAV8, and ~0.39 for full rAAV8) and medium to higher order retardation (~ 770 nm for empty rAAV9, ~1073 nm for full rAAV9, ~ 885 nm for empty rAAV8, and ~1342 nm for full

rAAV8) were also found in the crystals of capsids of serotypes AAV8 and AAV9 in **Fig. 3**. These retardation values suggest that the there is a significant difference in structural and electrical environment in different directions in the crystal/proteinaceous assembly and crystals/capsid proteins of all the serotypes have similar structural environment.

Because of their large size, individual macromolecular assemblies such as capsids in the crystals can be readily imaged with scanning electron microscopy (SEM). This can be used to reveal the mechanism of crystal growth, observe grain structure that confirms polycrystallinity/single crystal nature of crystals, and identify the presence of crystal defects without involving sophisticated diffraction methodologies. **Fig. 6** shows the SEM images of crystals for serotypes 5, 8, and 9 for both full (column 1 and 2) and empty (column 2 and 3) capsids at lower magnification (column 1 and 3) and at higher magnification (column 2 and 4).

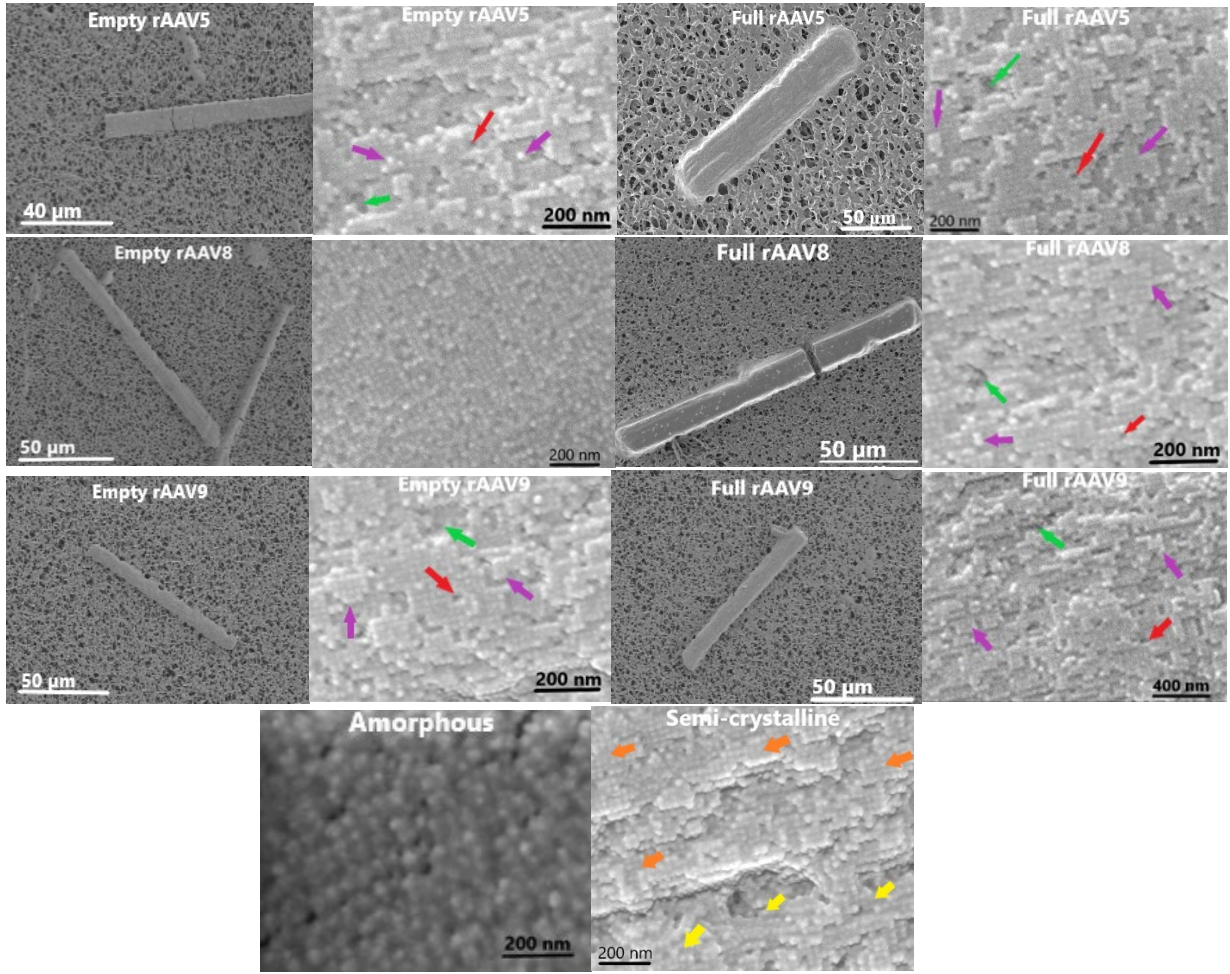

**Fig. 6:** Row 1: SEM images of crystals of empty rAAV5 at low magnification (column 1) and high magnification (column 2) and of full rAAV5 at low magnification (column 3) and high magnification (column 4). Row 2: SEM images of crystals of empty rAAV8 at low magnification (column 1) and high magnification (column 2) and of full rAAV8 at low magnification (column 3) and high magnification (column 4). Row 3: SEM images of crystals of empty rAAV9 at low magnification (column 1) and high magnification (column 2) and of full rAAV9 at low magnification (column3) and high magnification (column 4). Row 4: SEM image of a precipitate (left) and of a semi-crystalline solid (right). Experimental condition: 2.25% PEG8000, 0.9 M NaCl, pH 5.7 for empty rAAV5; 2% PEG, 0.6 M NaCl, pH 5.7 for full rAAV5; 3.4% PEG, 1 M NaCl, pH 5.7 for empty rAAV9; 3.5% PEG, 1.2 M NaCl, pH 5.7 for full rAAV9; 3.5% PEG, 0.6 M NaCl, pH 5.7 for empty rAAV8; 3% PEG, 0.3 M NaCl, pH 5.7 for full rAV8; 4.5% PEG,

0.65 M NaCl, pH 5.7, and empty rAAV9 for an amorphous material (precipitate); 2.5% PEG, 0.5 M NaCl, pH 5.7, and full rAAV9 for semi-crystalline solid.

These particles were obtained after crystallization in the common region of the crystallization as shown in our previous work.[32] As the **Fig. 6** higher magnification images suggest, for all serotypes, particles are made of small particles. Size of these small particles is same as that of capsids (~25 nm). Since there was no other material of that size in the system except capsids, these small particles are definitely capsids, which constituted the larger particles. These capsids are organised in a highly ordered manner, and this pattern of arrangement of capsids is repeated periodically over a long distance (long range order). Images also shows multiple layers of capsids representing each distinct plane. Presence of highly ordered capsids layers in the particle is a visual proof of the crystallinity of the large particles, because amorphous solids are characterized by highly disordered structure. As an example, **Fig. 6 row 4 (left)** shows a SEM image of an amorphous solid made of capsids. This shows no regular pattern of arrangement, except for a highly chaotic arrangement of capsids. Similarly, **Fig. 6 row 4 (right)** shows SEM image of a semi-crystalline particle, where capsids are arranged in a quasi-ordered fashion. Some regions in the particle possess highly ordered arrangement of capsids (orange arrow) as well as regions having chaotic (unordered) arrangement (yellow arrow). The highly ordered region resembles crystalline regions known as crystallites (orange arrow). Presence of crystallites within an amorphous matrix is a characteristic of a semi-crystalline material. Semi-crystalline particles are generally optically opaque due to the scattering of light from the crystalline-amorphous interface. Thus, the opaque solids formed in some of the crystallization conditions as observed in cross-polarized light microscopy and also reported in our previous work are semi-crystalline solid materials consisted of capsids.[33]

In each high magnification SEM image in **Fig. 6 column 2** and **4** in **rows 1**–**3**, it is observed that there are many small patches on the crystal surfaces. Within each patch, there is a highly ordered structured of capsids. Some of these patches are isolated and some are interconnected, which give rise to many "kink sites" as observed in the **Fig. 6**. This suggests that the growth of capsid crystal facets occurs by 2D nucleation (purple arrow) followed by growth of the 2D islets/patches via kink site attachment of capsids transported from the bulk or via the formation of many nuclei to spread the surface. As the images suggest, these 2D nuclei/islets may consist of one or more capsids. Because of its weak dependence on the supersaturation driving force, formation of 2D nucleation is feasible to occur at extremely low supersaturation level. This consistent with our previous finding that the capsid nucleation and growth in hanging-drop vapor diffusion system occurs at extremely low supersaturation.[42] Other growth mechanisms[52] such as spiral growth via screw dislocation as proposed in BCF model, normal growth as proposed by Wilson-Frenkel, and step flow growth models are absent in capsid crystal growth systems as suggested by the SEM images. Images also show that there are many incomplete layers and suggest that formation/stacking of new islets/layers begins before completion of the layer beneath, which is the characteristic of a random 2D nucleation-driven growth. SEM images of the crystals show that crystals possess many structural irregularities such as point defects (red arrow), line defects, volume defect (green arrow), and unidentified defects. Thus, by simply observing in SEM, it is possible to confirm their crystallinity, crystal structure, defects, growth mechanisms etc. because of the large size of the capsids. AFM is quite common in material science to observe the film structure and arrangement of atoms in the film or the lattice feature and defects using high-resolution topography image it provides, but AFM does not provide the higher depth of field, which is extremely important for crystals of massive structures such as capsids.

Elemental composition of the crystals is analyzed using Energy Dispersive X-ray (EDAX) to determine the distribution of different elements, which constitutes the capsids, and to determine the extent of the incorporation of foreign elements such as sodium, chlorine, and potassium during crystallization process. Since it is a bulk characterization technique and the high energy electron beam penetrates into the sample to interact with the atoms and release characteristic X-rays for the analysis, it's important to fix the energy of the electron beam. This analysis uses a beam energy/accelerating voltage of 15 kV, which ensures a penetration depth of 1 μm.[53] This penetration depth ensures that the interaction of the electron beam with the atoms is limited within the crystal only and information is not coming from the background membrane surface, which itself is composed of carbon and fluorine along with other elements. **Fig. 7a** shows the elemental mapping of the rAAV5 capsid crystals on the membrane surface. It shows that sodium and chlorine are almost absent in the crystal as shown by their relative distribution between crystal and membrane. Elements, which are present in the crystal in substantial amounts, are C, O, N, and P with relative abundance C followed by O followed by N followed by P as observed by their elemental distribution.

Salt crystals can be easily identified from the capsid crystals or the crystal of target material in the EDAX analysis by its elemental composition as shown in the **Fig. 7b** for NaCl crystal. It shows that Na and Cl are the two major elements in the crystal, and other element such as C, O, and N are present in extremely low quantity almost same as that in the background membrane. It seems that the salt crystal has exactly the same percentage of carbon as the background membrane, but the fact is that higher penetration depth (~ 1 µm) of electron beam (due to higher accelerating voltage) compared to the thickness of the salt crystal (< 1 µm) allows the interaction with carbon atoms in the membrane resulting in the presence of carbon in the elemental map of the salt crystal. It also seems that the relative abundance of O and N is more in salt than that in the membrane. This is probably because of the incorporation of some of the capsids or other protein impurities into the salt crystals during crystallization process.

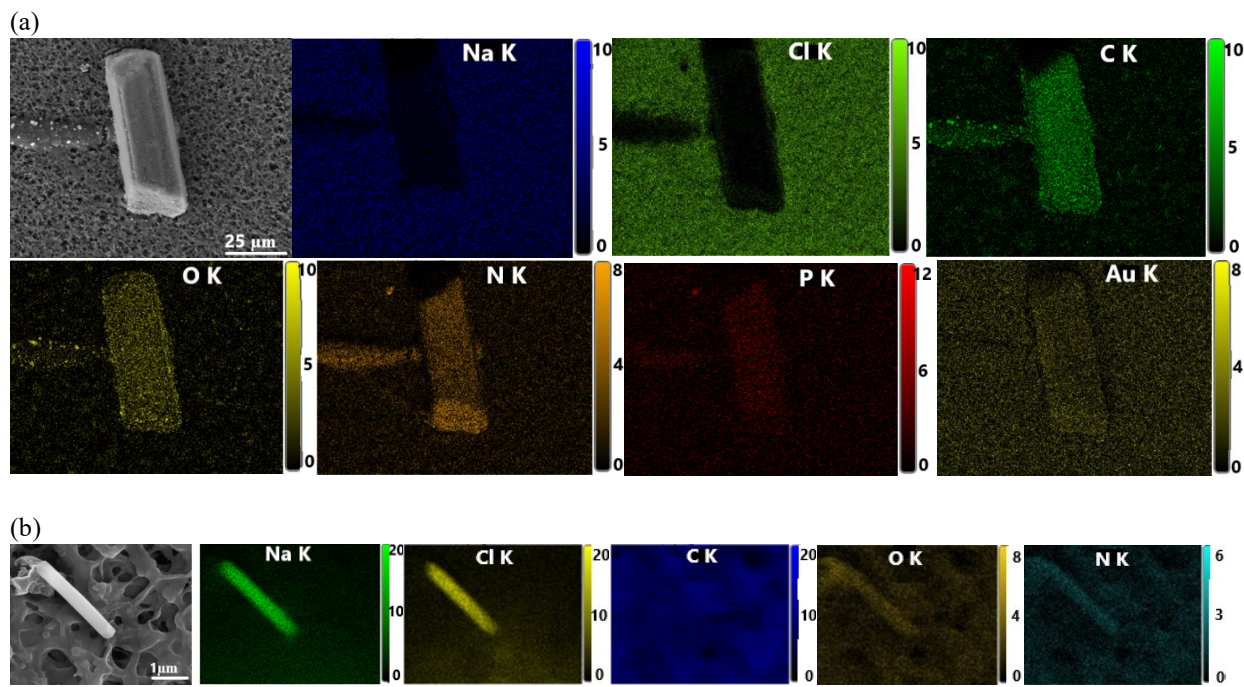

**Fig. 7:** EDAX analysis. (a) Elemental mapping of a crystal of full rAAV5 on membrane surface using EDAX. Row 1: SEM image of the crystal (column 1) and mapping of Na (column 2), Cl (column 3), and C (column 4); Row 2: mapping of O (column 1), N (column 2), P (column 3), and Au (column 4); Crystallization condition: 2.25% PEG, 1 M NaCl, pH 5.7 for rAAV5. (b) Elemental mapping of a salt crystal obtained in hanging drop vapor diffusion experiment. SEM image of the salt crystal (column 1) and mapping of Na (column 2), Cl (column 3), C (column 4), O (column 5), and N (column 6); Salt crystallization condition: 4% PEG, 5 M NaCl, and pH 5.7.

The line EDX was performed (**Fig. 8**) to obtain a quantitative measure of the elements in a crystal and capsid. The corresponding elemental composition of the crystals along the axial direction is shown in the **Table 1**. It shows that the atomic% of the undesired elements such as Na, K, and Cl is < 2 and the major element is C (atomic% > 55) followed by O (atomic% > 22), followed by N (atomic% > 10). This results in a ratio of $2.33 \pm 0.222 : 1 : 0.583 \pm 0.019$ for C:O:N. For all the serotypes, results remain roughly almost the same. However, the absence of any literature report on the elemental composition of capsids, EDX results could not be compared. Considering the fact that the C:O:N ratio varies depending on the amino acid sequence in the capsids, this result is consistent with that of 3.4:1 for C:N reported for bacteriophage proteins and of 3.6:1 for C:N reported from averaging of 2000 viral proteins.[54,55] Thus, the EDX analysis can be an alternative method to the conventional laborious and tedious protein sequencing analysis to obtain the elemental ratio/composition of a protein molecule or any biological molecules or proteinaceous assemblies such as capsids. Thus, EDAX can be useful in analyzing elemental composition of complex biological macromolecules such as capsids and in screening of crystals of target molecules and of crystallization conditions.

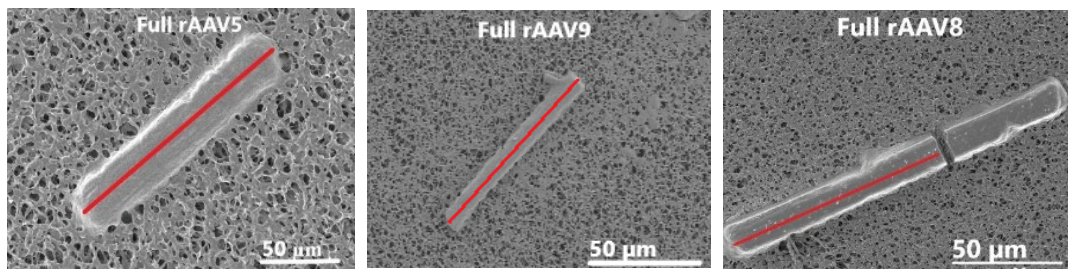

**Fig. 8:** Images showing the locations of the line-EDAX analysis on crystals for full rAAV5 (column 1), full rAAV8 (column 2), and full rAAV9 (column 3). Crystallization conditions: 2.5% PEG, 1.2 M NaCl, pH 5.7 for full rAAV5; 3.25% PEG, 1 M NaCl, pH 6.1 for rAAV8; 4% PEG, 1.1 M NaCl, pH 7.2 for rAAV9.

**Table 1:** Elemental compositon of crystals

| Elements | rAAV5 (atomic %) | rAAV9 (atomic %) | rAAV8 (atomic%) |
| --- | --- | --- | --- |
| C K | 54.11 ± 0.330 | 55.80 ± 1.270 | 54.65 ± 0.531 |
| N K | 13.53 ± 0.614 | 13.22 ± 1.203 | 13.75 ± 0.360 |
| O K | 23.20 ± 0.235 | 24.12 ± 0.510 | 23.87 ± 0.621 |
| Na K | 0.97 ± 0.085 | 1.09 ± 0.116 | 0.78 ± 0.151 |
| P K | 3.04 ± 0.270 | 3.17 ± 0.258 | 3.11 ± 0.165 |
| S K | 2.84 ± 0.357 | 2.71 ± 0.285 | 3.06 ± 0.148 |
| Cl K | 0.98 ± 0.049 | 0.86 ± 0.113 | 0.91 ± 0.287 |
| K K | 0.023 ± 0.006 | 0.03 ± 0.004 | 0.04 ± 0.005 |

Selected area electron diffraction (SAED) was performed for four different crystal forms (**Fig. 9**): rod (0.23 μm × 10 μm, empty rAAV8), cylinder (0.26 μm × 3 μm; full rAAV8), needle (0.24 μm × 50 μm for empty rAAV9, and 0.64 μm × 55 μm for full rAAV5, and thin plate (0.28 μm × 5 μm × 25 μm for full rAAV9, and 0.95 μm × 7.5 μm × 40 μm for empty rAAV5) to obtain more information about the capsid crystals. The small sample size is not amenable to powder X-ray diffraction analysis.[56,57] Crystals are small enough for analytical single crystal, low energy X-ray diffraction, that usually requires crystal ranging in size from 100 - 300 μm in all directions for an optimal diffraction pattern.[58,59] Advanced synchrotron radiation sources allow the use of smaller crystals, down to 6 μm, for single-crystal X-ray diffraction analysis, but access to synchrotron sources is limited and logistically inconvenient.[60]

Previous studies on single-crystal X-ray diffraction for AAV capsid crystals all used high energy Synchrotron source and diffraction grade crystals.[8,9,25] Stronger interaction between electron and the matter compared to X-rays produces diffraction pattern from small crystals. For an excellent cryo-electron diffraction pattern, sample thickness should be approximately 200–300 nm for an accelerating voltage of 200 keV, but even a 500 nm thick sample can produce a clear electron diffraction pattern with spots or rings of higher intensity for the same accelerating voltage.[61–63] Higher accelerating voltage allows the use of sample with even higher thickness for the optimal electron diffraction. For example, an accelerating voltage of 500 keV, even a sample thickness of 0.8–1 μm, can produce an excellent spot diffraction pattern.[63,64] However, most of the crystals grown in the hanging-drop vapor diffusion experiments possess size large/thick enough for even high energy electron beam to penetrate to generate electron diffraction pattern. But the crystal edges, which are around 230 nm thick in some cases, provide/offers "sweet spots" for the electron diffraction analysis and thus expected to produce an optimal spot diffraction pattern depending on the internal structure. But, in reality, these crystals with ~ 250 nm thickness did not produce any spot diffraction pattern. Instead, they produced the Kikuchi diffraction pattern as shown in **Fig. 9.**

Presence of clear well-defined Kikuchi lines in the electron diffraction pattern suggests that these crystals are single crystals. These crystals with approximately 250 nm thickness are still sufficiently thick enough to produce plural/multiple inelastic scattering (diffuse scattering) of electrons during theirs passage through the sample, leaving only a few transmitted electrons. This results in cones of scattering in 3D and produces lines, known as Kikuchi lines, on a 2D sensor when mapped.[65] Some of the Kikuchi patterns are shown in the **Fig. 9**. Many pairs of Kikuchi lines (KLs) as indicated by black double headed arrows with each pair consisted of a high-intensity line, known as KL excess, and a low-intensity line, known as KL defect, are formed. The KL defect and the KL excess, are shown in **Fig. 9 row 1** as an example by black and white arrows, respectively. Each Kikuchi pair, also known as Kikuchi band as shown by a double headed arrow, represents a set of diffraction planes with same Miller indices.

Both empty and full rAAV8 and full rAAV9 crystals form orthorhombic unit cell, while the empty rAAV9 forms the rhombohedral unit cell. The corresponding unit cell parameters for capsid crystals are a=35.6 nm, b=36.7 nm, c=37.5 nm for empty rAAV8; a=33.95 nm, b=31.93 nm, c=28.5 nm for full rAAV8; a=b=c=25.7 nm and α=β=γ=61° for empty rAAV9; a=39.3 nm, b=40.7 nm, c=41.87 nm for full rAAV9.

Miller indices are calculated from the interplanar spacing obtained from the distance between Kikuchi lines in each Kikuchi pair. **Fig. 9 row 1** shows the Miller indices for each unique diffraction plane/crystallographic plane responsible for the formation of each unique Kikuchi line pair or band. Based on the unit cell parameters, packing considerations suggests the presence of four complete viral capsids in the face-cantered orthorhombic unit cell for crystal of empty rAAV8, the presence of two complete viral capsids in the body-centered orthorhombic unit cell for the crystal of full rAAV8, the presence of one complete viral capsids in the primitive rhombohedral lattice for the crystal of empty rAAV9, and the presence of four complete viral capsids in the face-cantered orthorhombic unit cell for crystal of full rAAV9. Miller indices also suggest the presence of major planes such as {210},{120}, {102}, {121}, {111}, {100}, {001}, and {010} with {110}, {101}, and {011} having the highest number of capsids, which is the characteristic of a face-centered orthorhombic unit cell and the presence of {110}, {101}, {011}, {111} as the major planes with {111} containing highest number of atoms, which is the characteristic of a primitive rhombohedral lattice cell. Other prominent planes are {100}, {010}, {001}, {102}, and {112} etc., which are consistent with the orthorhombic unit cell. For all the crystals, higher interplanar spacing is consistent with the massive size of the macromolecular assemblies such as capsids. Absence of smaller interplanar spacing, e.g., ~ 0.282 nm associated with the prominent crystal planes {100} as seen in case of NaCl, rules out the formation of salt crystals.

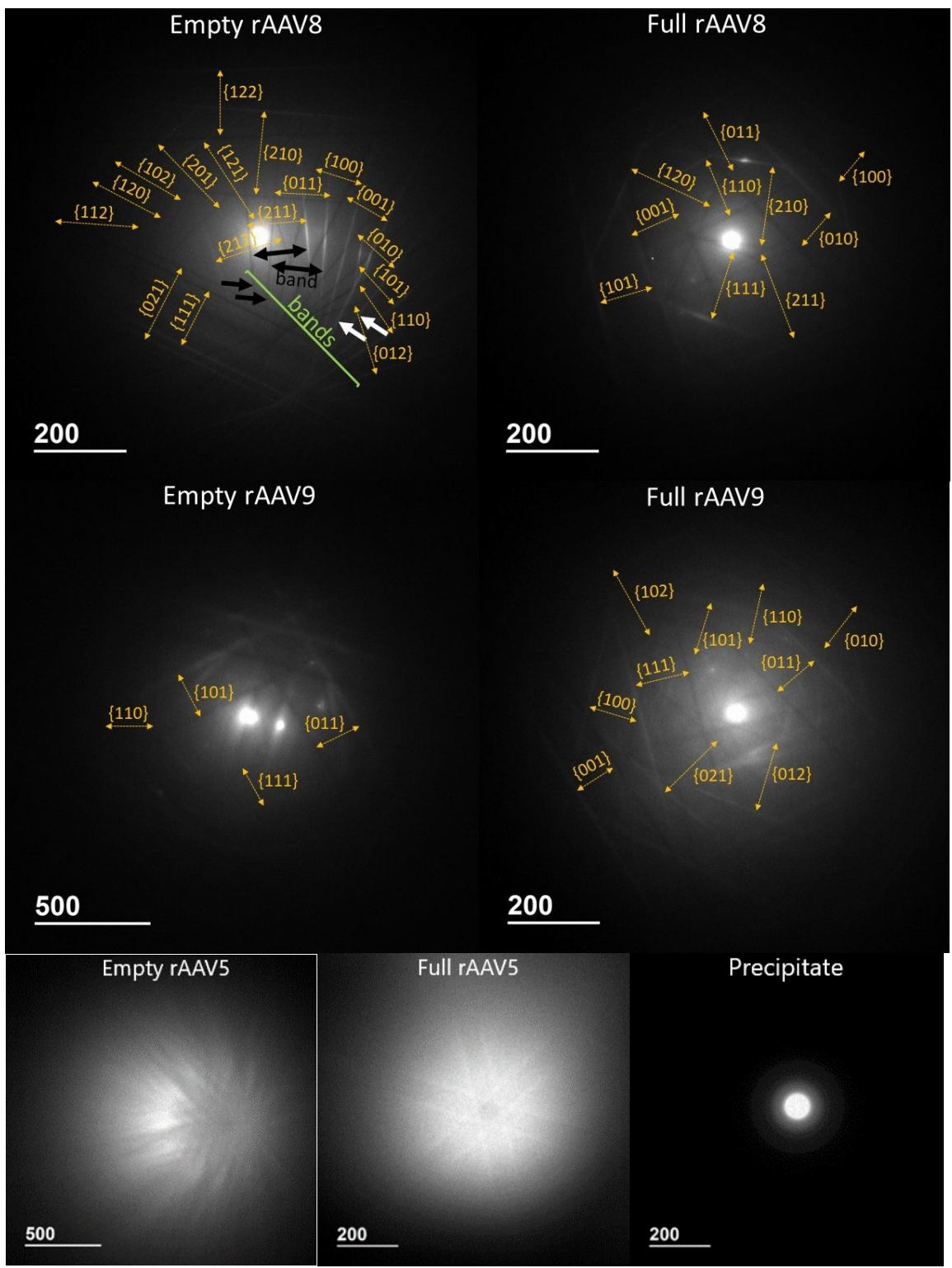

**Fig. 9:** Row 1: Electron diffraction pattern of empty rAAV8 crystal (column 1) and of full rAAV8 (column 2), Row 2: Electron diffraction pattern for empty rAAV9 (column 1) and full rAAV9 (column 2); Experimental condition: 3.5% PEG8000, 0.7 M NaCl, pH 7.2 for empty rAAV8; 3% PEG8000, 0.6 M NaCl, pH 6.1 for full rAAV8; 2.5% PEG8000, 1.2 M NaCl, pH 6.3 for empty rAAV9; 4% PEG8000, 1 M NaCl, pH 7.2 for full rAAV9. Row 3: Electron diffraction pattern of empty rAAV5 crystal (column 1), full rAAV5 crystal (column 2), and precipitate (column 3).

Experimental condition: 2.5% PEG8000, 0.65 M NaCl, pH 5.7 for empty rAAV5; 2.3% PEG8000, 0.75 M NaCl, pH 5.7 for full rAAV5; 5% PEG8000, 1.5 M NaCl, pH 7.2 for empty rAAV9 precipitate.

Sharp Kikuchi lines in diffraction pattern for some crystals (e.g., crystals in **Fig. 9 rows 1-2**) compared to the others (e.g., crystal in **Fig. 9 row 3)** indicate relatively higher thickness and more perfect crystal structure for the former than for the later. For the relatively thick crystals, the electron scatters so broadly that it diminishes the clarity of the Kikuchi diffraction pattern. However, too thick (~ 1 μm) crystals do not produce any diffraction pattern as observed in case of rAAV5 crystals. For very thick crystals, almost all the incident/penetrating electrons loose too much energy due to too many inelastic scatterings before reaching exit surface and this significantly reduces the coherent electron signal necessary for the formation of distinct interference pattern responsible for the Kikuchi pattern. Some of the crystals also produce prominent Kikuchi bands with the incident beam at the center as shown in **Fig. 9 row 3**. As usual the precipitate with the amorphous structure shows no diffraction pattern with a series of diffuse rings, which indicates a lack of long-range order in the material as shown in the **Fig. 9 row 3**. Thus, capsid crystals produce Kikuchi diffraction pattern for the crystals with thickness falling in the range ~300 nm to ~900 nm for an accelerating voltage of 400 keV for incident electron beam. For a low accelerating voltage of 200 keV, this range further decreases to ~220 nm to 600 nm (**Fig. 10**). Crystals with thickness higher than specified upper bound will not produce diffraction pattern. Capsid crystals with thickness below 200 nm are expected to produce an optimal spot diffraction pattern, which is essential for detailed structural studies using electron microscopy. This essentially suggests that for a clear spot diffraction pattern in a crystal of capsid, crystals must not incorporate more than 8 layers of capsids.

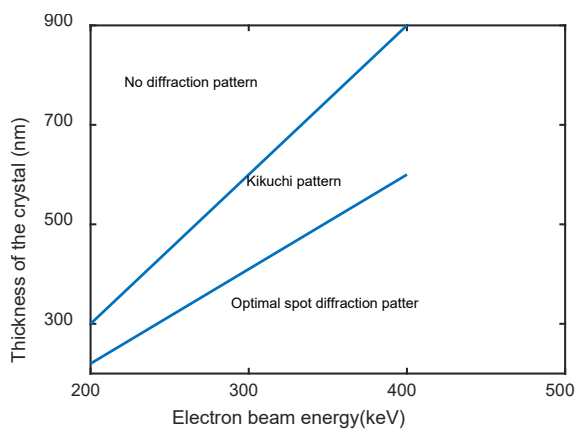

**Fig. 10:** Optimal conditions for obtaining diffraction pattern for capsid proteins in electron microscopy.

While the formation of excellent spot diffraction pattern for a sample of approximate thickness 200–800 nm is guaranteed for inorganic or small organic molecules or small proteins such as lysozyme, the formation of Kikuchi diffraction pattern even for crystals of thickness approximately 250 nm is probably the characteristic of crystals of the massive macromolecular assemblies such as capsids. This is because their heavy molecular structure with zillions of low atomic number atoms such as hydrogen, and carbon with numerous covalent bonds and readily accessible excited states facilitates multiple inelastic electron scattering (diffuse scattering) events, necessary for the formation of strong Kikuchi pattern, by vibrational, rotational or electronic excitation, which is not easily possible in most of the inorganic, and organic crystals and may be feasible to a limited extent in small protein molecules. This causes the capsid crystals to produce significant number of inelastic scattering events and resulting Kikuchi pattern. Thus, obtaining a clear spot diffraction pattern for capsid crystals needs much thinner sample, probably < 200 nm, and higher accelerating voltage (> 400 keV).

**4 Conclusion**

This work describes the characterization of AAV capsid crystals (AAV serotypes 5, 8, and 9) as model complex biological macromolecular assemblies. As discussed in this work, cross-polarized light microscopy can be used to identify capsid crystals in a mixture of salt crystals, hair fibers, precipitate materials, dense solid phase, and opaque crystals, which are very common in the crystallization of complex biological macromolecular assemblies and are not seen in the crystallization system involving small protein molecules, and thus allows the rapid screening of their crystallization conditions. It is possible because cross-polarized light microscopy can distinguish the anisotropic crystals of proteinaceous assemblies from the crystals of Kosmotropic salts such as NaCl and Chaotropic salts such as KCl, $MgCl_2$, and $CaCl_2$, as these salts mostly form optically isotropic crystals and from the non-birefringent opaque crystals and dense materials. It is found that there is a significant difference in structural and electrical environment in the crystal/proteinaceous assembly in the different directions and crystals/capsid proteins of all the serotypes have similar structural environment as suggested by their birefringence and optical retardation value.

As the analysis suggests, SEM imaging can be a potential alternative to the AFM, which is frequently used in the study of surface or growth of crystals made of small molecules or atoms. The not-so-common SEM imaging can be exploited either to confirm the crystallinity or to understand the mechanism of crystallization of massive macromolecular assemblies such as capsids, just by observing their arrangement at higher magnification. It is found that the growth of the crystal facets occurs by the random 2D nucleation forming islets, which is consisted of one or more capsids, followed the growth of the 2D islets via the attachment of capsids at the kink sites and/or via the formation of more 2D nuclei to spread the already formed islets of 2D nuclei. Other growth mechanisms prominent at the low supersaturation environment and for the small inorganic and organic molecules and atoms such as spiral growth as suggested by BCF model, step growth, and the normal growth as proposed by the Wilson-Frankel are not preferred by capsids. SEM imaging finds that the capsids form crystallites and resulting semi-crystalline solid material, which is identified as an optically opaque material in the cross-polarized light microscope. It is also found that the capsid crystals inherit many structural irregularities in the form of lattice defects such as volume defects, point defects, and many unrecognized defects.

As suggested by the EDAX analysis, capsid crystals contain C, O, and N as major elements in the ratio of $2.33 \pm 0.222 : 1 : 0.583 \pm 0.019$. This ratio remains nearly the same for all serotypes of rAAV. It can be concluded that the EDX analysis, which is regularly used in material science to determine the film/crystal composition, can be used to obtain the elemental ratio/composition of a protein molecule or any biological molecules or proteinaceous assemblies such as capsids as an alternative to the conventional protein sequencing method commonly used in the biology.

It is found that the capsid crystals generate a Kikuchi diffraction pattern for crystals of thickness falling in the rage ~300 nm to ~900 nm (at 400 keV accelerating voltage) and ~250 nm to 600 nm, which generally produce an excellent spot diffraction pattern for crystals of small inorganic or organic molecules. For an optimal spot diffraction pattern for capsid crystals, sample thickness is required to be $\lesssim$ 200 nm, which is equivalent to a crystal with at least one dimension consisting of at most 8 layers of capsids. Formation of a Kikuchi pattern for crystals of capsids of thicknesses, which are expected to produce spot diffraction pattern, seems to be the characteristics of complex biological macromolecular assemblies. This is possible because the massive proteinaceous assemblies such as capsids possess very large numbers of low atomic number atoms such as carbon, hydrogen, and oxygen along with numerous covalent bonds provides readily accessible excited states (by vibration, rotational, and electronic excitation) necessary for the plural inelastic electron scattering (diffuse scattering) events responsible for the formation of a Kikuchi diffraction pattern. It is found from the analysis of the Kikuchi diffraction pattern that capsids form face-centered orthorhombic unit cell with four complete capsids, body-centered orthorhombic unit cells with two complete capsids, and rhombohedral primitive lattice with one complete capsid. Thus, the characterization methods presented in this work will be useful in studying the crystals of biological molecules and other complex macromolecular assemblies and designing new materials with specific properties.

**Author contributions:**

V.B. conceptualized this work, designed and conducted the experiments, developed models, wrote code, performed simulation, analyzed the data, and wrote the initial draft. R.D.B. conceptualized this work, supervised the work/project, edited the manuscript, and acquired the funds to support the project. R.M.K., P.W.B., and S.L.S. edited the manuscript, supervised the work/ project, and acquired the funds to support the project. A.J.S. and J.M.W. edited the manuscript, supervised the project, and acquired the funds to support the project.

**Acknowledgements:**
Funding is acknowledged from the MLSC, Sanofi, Sartorius, Artemis, and US FDA (75F40121C00131).

**References**

[1] T. W. Hayton, S. M. Humphrey, B. M. Cossairt, R. L. Brutchey, *Inorg Chem* **2023**, *62*, 13165.

[2] A. S. Panwar, A. Singh, S. Sehgal, *Mater Today Proc* **2020**, *28*, 1932.

[3] "Characterization," can be found under https://dmse.mit.edu/research-impact/materials-research-type/characterization/, **n.d.**

[4] K. L. Chen, H. T. Liu, J. H. Yu, Y. H. Tung, Y. S. Chou, C. C. Yang, J. S. Wang, J. L. Shen, K. C. Chiu, *Sci Rep* **2018**, *8*, 16740.

[5] "Pharmaceutical testing & raw material characterization," can be found under https://www.impactanalytical.com/pharmaceutical-testing-and-raw-material-characterization/, **n.d.**

[6] M. Mitchell, H. J. Nam, A. Carter, A. McCall, C. Rence, A. Bennett, B. Gurda, R. McKenna, M. Porter, Y. Sakai, B. J. Byrne, N. Muzyczka, G. Aslanidi, S. Zolotukhin, M. Agbandje McKennaa, *Acta Crystallogr Sect F Struct Biol cryst Commun* **2009**, *65*, 715.

[7] E. Trilisky, R. Gillespie, T. D. Osslund, S. Vunnum, *Biotechnol. Prog.* **2011**, *27*, 1054.

[8] Q. Xie, H. M. Ongley, J. Hare, M. S. Chapmana, *Acta Crystallogr Sect F Struct Biol Cryst Commun* **2008**, *64*, 1074.

[9] Q. Xie, W. Bu, S. Bhatia, J. Hare, T. Somasundaram, A. Azzi, M. S. Chapman, *Proc Natl Acad Sci U S A* **2002**, *99*, 10405.

[10] M. Scimeca, S. Bischetti, H. K. Lamsira, R. Bonfiglio, E. Bonanno, *European Journal of Histochemistry* **2018**, *62*, 2841.

[11] B. Kascakova, A. Koutska, M. Burdova, P. Havlıckova, I. K. Smatanov, *FEBS Open Bio* **2024**.

[12] M. Mustafa, H. C. Kim, H. D. Yang, K. H. Choi, *Journal of Materials Science: Materials in Electronics* **2013**, *24*, 4321.

[13] J. Kang, N. Shin, D. Y. Jang, V. M. Prabhu, D. Y. Yoon, *J Am Chem Soc* **2008**, *130*, 12273.

[14] K. Takaba, S. M. Yonekura, I. Inoue, K. Tono, T. Hamaguchi, K. Kawakami, H. Naitow, T. Ishikawa, M. Yabashi, K. Yonekura, *Nat Chem* **2023**, *15*, 491.

[15] A. Joseph, M. M. Xavier, G. Żyła, P. R. Nair, A. S. Padmanabhan, S. Mathew, *RSC Adv.* **2017**, *7*, 16623.

[16] "Material characterization," can be found under https://www.cambrex.com/analytical-services/analytical-testing/material-characterization/, **n.d.**

[17] Y. Yamada, K. Kihira, Iwata. M., S. Takahashi, K. Inaka, H. Tanaka, I. Yoshizaki, in *Handbook of Space Pharmaceuticals*, Springer Nature, **2022**, pp. 887–912.

[18] F. Qiao, T. A. Binknowski, I. Broughan, W. Chen, A. Natarajan, G. E. Schiltz, K. A. Scheidt, W. F. Anderson, R. Bergan, *bioRxiv* **2024**.

[19] G. Holdgate, S. Geschwindner, A. Breeze, G. Davies, N. Colclough, D. Temesi, L.

Ward, *Methods Mol Biol .* **2013**, *1008*, 327.

[20] A. A. Kermani, *FEBS J* **2020**, *288*, 5788.

[21] Xi. H. Tang, J. J. Liu, Y. Zhang, X. Z. Wang, *J Cryst Growth* **2018**, *498*, 186.

[22] H. Xu, H. Lebrette, T. Yang, V. Srinivas, S. Hovmöller, M. Högbom, X. Zou, *Structure* **2018**, *26*, 667.

[23] W. Singer, H. Rubinsztein-Dunlop, U. Gibson, *Opt Express* **2004**, *12*, 6440.

[24] A. A. Echalier, R. L. Glazer, V. Fülöpa, M. A. Geday, *Acta Crystallogr D structural biology* **2004**, *60*, 696.

[25] T. F. Lerch, Q. Xie, H. M. Ongley, J. Hare, M. S. Chapman, *Acta Crystallogr Sect F Struct Biol Cryst Commun* **2009**, *65*, 177.

[26] Q. Xie, J. Hare, J. Turnigan, M. S. Chapman, *J Virol Methods* **2004**, *122*, 17.

[27] J. M. Spear, A. J. Noble, Q. Xie, D. R. Sousa, M. S. Chapman, S. M. Stagg, *J Struct Biol.* **2015**, *192*, 196.

[28] S. M. Stagg, C. Yoshioka, O. Davulcu, M. S. Chapman, *Chem Rev* **2022**, *122*, 14018.

[29] V. Rayaprolu, S. Kruse, R. Kant, R. Kant, B. Venkatakrishnan, N. Movahed, D. Brooke, B. Lins, A. Bennett, T. Potter, R. McKenna, M. Agbandje-McKenna, B. Bothner, *J Virol* **2013**, *87*, 13150.

[30] T. P. Wörner, A. Bennett, S. Habka, J. Snijder, O. Friese, T. Powers, M. Agbandje-McKenna, A. J. R. Heck, *Nat Commun* **2021**, *12*, 1642.

[31] D. Montiel-Garcia, N. Santoyo-Rivera, P. Ho, M. Carrillo-Tripp, J. E. Johnson, V. S. Reddy, *Nucleic Acids Res* **2020**, *49*, D809.

[32] V. Bal, J. M. Wolfrum, P. W. Barone, S. L. Springs, A. J. Sinskey, R. M. Kotin, R. D. Braatz, *Selective Enrichment of Full AAV Capsids*, **2024**.

[33] V. Bal, J. M. Wolfrum, P. W. Barone, S. L. Springs, A. J. Sinskey, R. M. Kotin, R. D. Braatz, **2025**, DOI https://doi.org/10.48550/arXiv.2501.18104.

[34] W. D. Nesse, *Introduction to Optical Minerology.*, Oxford University Press, **2004**.

[35] M. D. , M. E. G. and D. Tasa. Dyar, *Mineralogical Society of America* **2008**.

[36] J. Kalb, F. Spaepen, M. Wuttig, *Appl. Phys. Lett.* **2004**, *84*, 5240.

[37] Yu. G. Kuznetsov, A. J. Malkin, A. McPherson, *J Cryst Growth* **1999**, *196*, 489.

[38] M. Golub, A. Graja, K. Jóźwiak, *Synth Met* **144AD**, *22*, 201.

[39] X. Mu, C. Gillman, C. Nguyen, T. Gonen, *Annu Rev Biochem.* **2021**, *90*, 431.

[40] H.-J. Nam, B. L. Gurda, R. McKenna, M. Potter, B. Byrne, M. Salganik, N. Muzyczka, M. Agbandje-McKenna, *J Virol* **2011**, *85*, 11791.

[41] E. B. Miller, B. G. Whitaker, L. Govindasamy, R. McKenna, S. Zolotukhin, N. Muzyczka, M. A. McKenna, *Acta Crystallogr Sect F Struct Biol Cryst Commun* **2006**, *62*, 1271.

[42] V. Bal, M. S. Hong, J. M. Wolfrum, P. W. Barone, S. L. Springs, A. J. Sinskey, R. M. Kotin, R. D. Braatz, *Cryst Growth Des* **2025**, *25*, 3687.

[43] A. M. Gruntman, L. Su, Q. Su, G. Gao, C. Mueller, T. R. Flotte, *Hum Gene Ther Methods* **2015**, *26*, 71.

[44] B. Lins-Austin, S. Patel, M. Mietzsch, D. Brooke, A. Bennett, B. Venkatakrishnan, K. Van Vliet, A. N. Smith, J. R. Long, R. McKenna, M. Potter, B. Byrne, S. L. Boye, B. Bothner, R. Heilbronn, M. Agbandje-McKenna, *Viruses* **2020**, *12*, 1.

[45] H. P. Stevenson, G. Lin, C. O. Barnes, I. Sutkeviciute, T. Krzysiak, S. C. Weiss, S. Reynolds, Y. Wu, V. Nagarajan, A. M. Makhov, R. Lawrence, E. Lamm, L. Clark, T. J. Gardella, B. G. Hogue, C. M. Ogata, J. Ahn, A. M. Gronenborn, J. F. Conway, J. P. Vilardaga, A. E. Cohen, G. Calero, *Acta Crystallogr D Struct Biol* **2016**, *72*, 603.

[46] N. Davidson, J. Hillier, *J Appl Phys* **1947**, *18*, 499.

[47] W. D. Nesse, *Introduction to Optical Minerology*, Oxford University Press, New York, NY, **2004**.

[48] F. D. Bloss, *Crystallography and Crystal Chemistry*, Hold, Rinehart, Winston,Inc., New York, NY, **2012**.

[49]  and D. Tasa. Dyar, M. D., M. E. Gunter, *Mineralogy and Optical Mineralogy*, Mineralogical Society Of America, Chantilly, VA, **2008**.

[50] P. J. Polowsky, G. F. Tansman, P. S. Kindstedt, J. M. Hughes, *J Dairy Sci* **2018**, *101*, 7714.

[51] "Reflected light DIC microscopy," can be found under https://www.microscopyu.com/techniques/dic/reflected-light-dic-microscopy, **n.d.**

[52] H. M. Cuppen, Theory and Simulations of Crystal Growth. Fundamental Steps in Morphology Prediction:, Radboud University, **2005**.

[53] "Optimizing spatial resolution for EDS analysis," can be found under https://www.edax.com/-/media/ametekedax/files/resources/tips_tricks/optimizingspatialresolutionforedsanalysis.pdf, **n.d.**

[54] U. M, "Ratio of carbon and nitrogen in amino acid (discovered by analysis of primary sequence of more than 2000 viral proteins)," can be found under https://bionumbers.hms.harvard.edu/bionumber.aspx?s=n&v=7&id=113053#:~:text=Ratio%20of%20carbon%20and%20nitrogen%20in%20amino%20acid%20(discovered%20by,more%20than%202000%20viral%20proteins)&text=Comments-,P.,and%20nitrogen%20poor%20than%20DNA.%22, **n.d.**

[55] L. F. Jover, T. C. Effler, A. Buchan, S. W. Wilhelm, J. S. Weitz, *Nat Rev Microbiol* **2014**, *12*, 519.

[56] "X-Ray diffraction laboratory," can be found under https://cms.eas.ualberta.ca/xrd/, **n.d.**

[57] "X-Ray diffraction (XRD) analysis," can be found under https://sprinttesting.com/x-ray-diffraction-xrd.html, **n.d.**

[58] "Single-crystal X-ray diffraction," **n.d.**

[59] "Growing quality crystals," can be found under https://chemistry.mit.edu/facilities-and-centers/x-ray-diffraction-facility/growing-quality-crystals/, **n.d.**

[60] J. J. Pluth, J. V. Smith, D. Y. Pushcharovsky, R. W. Broach, *Earth, Atmospheric, and Planetary Sciences* **94AD**, *23*, 12263.

[61] Z. Huang, M. Ge, F. Carraro, C. Doonan, P. Falcaro, X. Zou, *Faraday Discuss.* **2021**, *225*, 118.

[62] J. Rodriguez, M. I. Ivanova, M. R. Sawaya, D. Cascio, F. E. Reyes, D. Shi, S. Sangwan, E. L. Guenther, L. M. Johnson, M. Zhang, L. Jiang, M. A. Arbing, B. L. Nannenga, J. Hattne, J. Whitelegge, A. S. Brewster, M. Messerschmidt, S. Boutet, N. K. Sauter, T. Gonen, D. S. Eisenberg, *Nature* **2015**, *525*, 486.

[63] R. Uyeda, M. Nonoyama, *Jpn J Appl Phys* **1968**, *7*, 200.

[64] Y. Liao, "Overall electron diffraction and Kikuchi lines depending on TEM sample thickness - practical electron microscopy and database," **n.d.**

[65] B. Fultz, J. Howe, in *Transmission Electron Microscopy and Diffractometry of Materials*, **2013**, pp. 289–348.